\documentclass[aps,pra,twocolumn,showpacs,superscriptaddress,10pt]{revtex4-1}

\usepackage{amsmath}
\usepackage{amsfonts}
\usepackage{amssymb}
\usepackage{graphicx}
\usepackage[export]{adjustbox}

\usepackage{xcolor}
\usepackage{color}
\usepackage{mathtools}
\usepackage{mathrsfs}
\usepackage[normalem]{ulem}

\global\long\def\ket#1{\left| #1\right\rangle }
 \global\long\def\bra#1{\left\langle #1 \right|}
 \global\long\def\av#1{\left\langle #1 \right\rangle }
  \global\long\def\abs#1{\left| #1 \right| }
  \definecolor{dgreen}{rgb}{0.0, 0.5, 0.0}

\begin{document}
%
%
\title{Spectrum, Landau-Zener theory  and driven-dissipative dynamics \\
of  a staircase of photons}
\author{J. Marino}
\email{jamirmarino@fas.harvard.edu} 
\affiliation{Department of Physics, Harvard University, Cambridge MA 02138, United States\\
Department of Quantum Matter Physics, University of Geneva, 1211, Geneve, Switzerland}

\author{Y. E. Shchadilova} 
\affiliation{Department of Physics, Harvard University, Cambridge, Massachusetts 02138, USA}

\author{M. Schleier-Smith}
\affiliation{Department of Physics, Stanford University, Stanford, CA 94305, USA}

\author{E. A. Demler} 
\affiliation{Department of Physics, Harvard University, Cambridge, Massachusetts 02138, USA}

\begin{abstract}
%
We study  the production of photons in a model of three  
  bosonic atomic modes non-linearly coupled to a cavity mode. 
 %
%
In absence of external driving and dissipation, the  energy levels at different photon numbers assemble into the steps of an energy staircase which can be employed as guidance for preparing multi-photon states.
We consider adiabatic photon production,  driving the system through a sequence of Landau-Zener transitions in the presence of external coherent light pumping.  We also analyse the non-equilibrium dynamics of the system resulting from the  competition of the sudden switch of coherent photon pumping and cavity photon losses, and
we find that the system approaches a plateau with a given number of photons, which becomes metastable upon increasing the rate of photon pumping.  We discuss the sensitivity of the time scales for the onset of this metastable behaviour to system parameters and predict the  value of photons attained, solving the driven-dissipative dynamics including three-body correlations between light and matter degrees of freedom.

\end{abstract}


\date{\today}
\maketitle

\section{Introduction} 
 The last ten years have witnessed swift progress in quantum optics platforms where  light and matter are strongly coupled and can be employed to engineer a variety of quantum phenomena:
examples range from Bose-Einstein condensates coupled to optical cavity photons, where the Dicke transition is engineered~\cite{hemm, ess}, to the recent demonstration of  supersolids~\cite{leonard} and phases with competing order parameters in a condensate trapped at the intersection of two optical cavities~\cite{leonard2, leonard3, gopal, piazza}.  
Most of these models realise scenarios where matter and light collectively interact as in the case of superradiant phase transitions or in 'Dicke-Hubbard'  systems characterised by critical points separating a superfluid from a Mott insulating phase~\cite{hemm2}.  Strong light-matter coupling regimes have also enabled the preparation of highly squeezed states of atomic ensembles by quantum non-demolition measurements \cite{host, bohnet}, photon-mediated spin interactions \cite{lero,hosten2016quantum} in optical cavities, photon blockade effects~\cite{hartman, hamsen, cirac, rein, ima} or  non-classical light in cavity optomechanics platforms~\cite{brooks}.  Recent experiments have extended photon-mediated interactions to optical-clock atoms \cite{norcia}, to spin-1 atoms \cite{Zhiqiang:17,davis2018photon}, and to multi-mode cavities \cite{vaidya2018tunable}, which enable further advances in quantum metrology \cite{masson} and quantum simulation.


The control and the preparation of multi-photon states is of paramount importance for a progress towards a many-body physics of coupled light and matter in these platforms.
Single- and multi-photon preparation has a long history in cavity QED~\cite{brattke, bertet, uren, sayrin, cooper, schuster, hafezi}, including photon generation in high quality cavities~\cite{varcoe}, the control of single-photon states emitted by polaritons~\cite{stan}, as well as the conversion of collective atomic excitations into single photonic states within optical resonators~\cite{vuletic}, encompassing quantum homodyne tomography~\cite{lv01} and preparation of photon states in the driven dissipative dynamics of cavity arrays~\cite{tomadin, leb}.

 In this work we consider a novel cavity-QED platform composed of collective atomic degrees of freedom strongly non-linearly coupled to a cavity photon: this system can be employed to engineer multi-photon states out of an empty cavity via adiabatic as well as with far-from-equilibrium driving protocols.
Specifically, we study an effective model of two bosonic atomic modes interacting with a photon; the model results from a two-photon resonant process occurring in a cavity hosting an ensemble of spin-1 atoms Rabi-coupled to the cavity mode. %
In equilibrium conditions, the eigenstates of this system compose a `staircase' structure: each level, or `step', is characterised by a different photon number which is a conserved quantity in undriven conditions. 
In the presence of weak coherent photonic pumping, an adiabatic variation in time of the energy levels of the atomic degrees of freedom allows for climbing the staircase, transiting across a sequence of level crossings, and thus preparing a desired number of photons out of an initially empty cavity.
Complementarily, we also consider the far-from-equilibrium dynamical preparation of photons in the system,  suddenly switching the coherent  photon pumping as well as including natural sources of dissipation, such as incoherent cavity-photon losses.
We highlight the formation of a metastable steady state in the late time driven-dissipative dynamics of photons, and discuss the dependence of its life-time on system parameters, solving the dynamics with the inclusion of three-body correlations between light and atomic degrees of freedom.
\\

\section{The model } 
\begin{figure}[t!]
\includegraphics[width=8cm, frame]{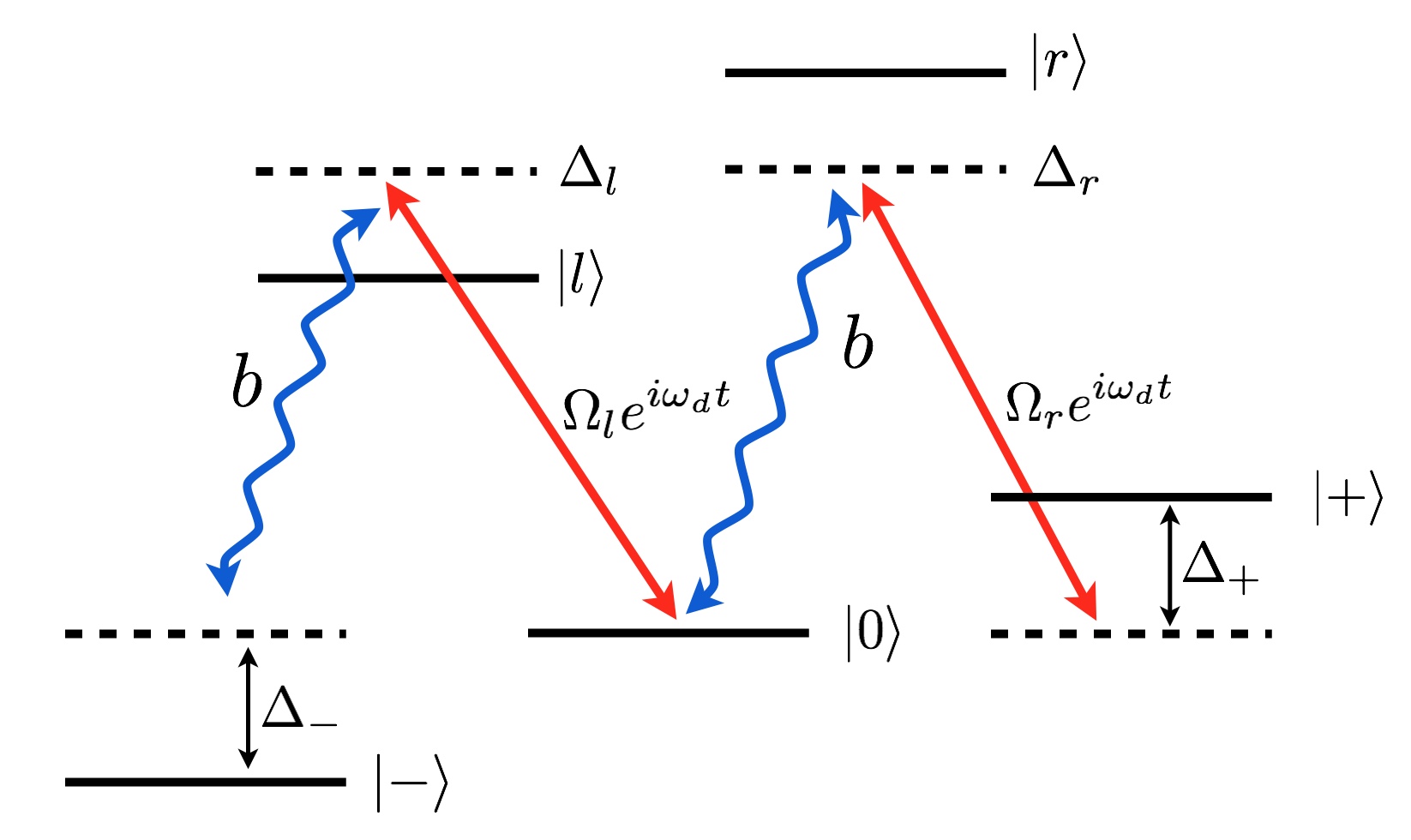}
\caption{Schematics of the optical transitions considered in this work: a two-photon process transferring two atoms from the level $|0\rangle$ (with macroscopic occupation $\mathcal{N}_0$) to the levels $|+\rangle$ and $|-\rangle$ becomes resonant when the two detunings $\Delta_+$ and  $\Delta_-$ compensate each other: $\Delta_+=-\Delta_-$. The transition is assisted by  two lasers of frequency $\omega_d$ (red straight lines) and by the two cavity photons (blue wiggled lines). }
\label{fig1}
\end{figure}

We consider an optical cavity supporting a photonic mode ($b$ in Fig.~\ref{fig1}) of frequency $\omega_b$, Rabi coupled via the interaction coupling $g$ to the atoms.
The three ground energy levels, e.g. Zeeman states in an atom of hyperfine spin $F=1$ (\cite{Zhiqiang:17,masson,davis2018photon}), are denoted with $|+\rangle$, $|-\rangle$, $|0\rangle$, and the first two are detuned upwards and downwards,  $\Delta_+>0$ and $\Delta_-<0$  with respect to the latter, which is assumed to have a macroscopic occupation, $\mathcal{N}_0\gg1$. 
A couple of external lasers of frequency $\omega_d$ can assist transitions from the two levels $|0\rangle $ and $|+\rangle$ to two auxiliary levels $|r\rangle$ and $|l\rangle$ respectively, with amplitudes $\Omega_{l,r}$. The lasers are far detuned from the transitions to the auxiliary levels by $\Delta_{l/r}$.
Single-atom transitions assisted by the laser $r$ (or $l$) and by a cavity photon, transferring population from the states $|\pm\rangle$ to $|0\rangle$ (through levels $|r\rangle$ and $|l\rangle$) are off-resonant by an amount $\delta_{\pm}=\omega_d-\omega_b-\Delta_\pm$. However, transitions from the atomic state $|0,0\rangle$ to $|+,-\rangle$, involving two atoms and assisted by a virtual photon emitted into the cavity and then rescattered,  are resonant if the two detunings compensate each other,  $\Delta_+=-\Delta_-$ (a schematic of the energy levels and of the transitions is provided in Fig.~\ref{fig1}).
This resonant transition is pivotal for the realization of the photonic staircase at the core of this work, and the associated effective Hamiltonian reads 
\begin{equation}\label{eq:ham}
\begin{split}
 H &= \omega  n_b + \epsilon_+  a_+^\dag  a_+ + \epsilon_-  a_-^\dag  a_- + \\
&+\lambda (b~b^\dag+b^\dag b) ( a_+  a_- +  a_+^\dag  a_-^\dag).
\end{split}
\end{equation}
The last term embodies the photon-assisted resonance process changing simultaneously the population of the two atomic levels $|\pm\rangle$. In Eq.~\eqref{eq:ham} we have reabsorbed the large occupation, $\mathcal{N}_0$, of the $|0\rangle$ level in the coupling $\lambda$; the mode  $|0\rangle$ can therefore be treated classically, while we assume that the occupation of the levels $|\pm\rangle$ remains small.
The frequency $\omega$ stands for the cavity mode frequency relative to the frequency of the lasers.

This derivation follows the lines of Ref.~\cite{borregard17}, considering 
 $N$ three-level atoms Rabi coupled to a single mode optical cavity of frequency $\omega_b$. Each atom has an internal structure consisting of the three states $\ket{+}$, $\ket{0}$, $\ket{-}$  (ordered with decreasing energy),  and the two states $|\pm\rangle$ have a different Rabi coupling constant, $g_{\pm}$, with the photonic cavity mode. With the energies of detunings, $\Delta_{l/r}$ and $\delta$, larger than all the other energy scales involved in the system, one can microscopically derive, via adiabatic elimination, the hamiltonian~\eqref{eq:ham}. We find that 
 \begin{equation}\begin{split}\label{quanti}
 \epsilon_+&={\Omega_r^2 \mathcal{N}_0}/{\Delta_r}+q,\quad
 \epsilon_-={\Omega_l^2\mathcal{N}_0}/{\Delta_l}+q,\\
 &\lambda=g_+ g_-  \Omega_l \Omega_r\mathcal{N}_0/{ (2 \delta \Delta_l \Delta_r)},
\end{split} \end{equation}
 %
 where $q=\Delta_++\Delta_-$ accounts for the quadratic Zeeman shift, and $\delta\simeq\delta_{\pm}$, assuming that the difference between $\delta_{\pm}$ (controlled by $q$) is smaller than their average.
Following this procedure, one finds that  $a_{+}$ and $a_-$ are collective operators summing over all the single-particle excitations of the $N$ atoms, and they therefore have bosonic commutation relations.\\

\section{A staircase of photons } The hamiltonian ~\eqref{eq:ham} conserves the number of photons  in the system, $n_b\equiv b^\dag b$,  allowing for diagonalization in sectors of the Hilbert space with fixed number of photons, $n$.  We describe, in each of these sectors, the lowest energy state using the variational ansatz state 
\begin{eqnarray}\label{eq:WF}
\ket{\Psi(\phi_n)} = e^{\phi_n  a_+^\dag a_-^\dag - \textsl{h.c.}} \ket{\psi} \equiv e^{S_n}\ket{\psi},
\end{eqnarray}
where $\ket{\psi}$ is the vacuum state simultaneously annihilated by $a_\pm$. The  unitary map $e^{ S}$ transforms  the operators $ a_+$ and $ a_-$ as 
\begin{equation}\begin{split}\label{eq:transfs}
 \tilde{a}_+=e^{- S_n}  a_+  e^{ S_n} =& \cosh(\phi_n) a_+ +  \sinh(\phi_n) a_-^\dag, \\
 \tilde{a}_-= e^{- S_n}  a_-   e^{ S_n} =& \sinh(\phi_n) a_+^\dag  +  \cosh(\phi_n) a_-,
\end{split}\end{equation}
yielding the following expectation values 
\begin{equation}
\begin{split}
\bra{\Psi(\phi_n)} a_+ a_-\ket{\Psi(\phi_n)} &= \cosh(\phi_n) \sinh(\phi_n), \\
\bra{\Psi(\phi_n)} a_+^\dag a_+\ket{\Psi(\phi_n)} &= \sinh^2(\phi_n).
\end{split}
\end{equation}
The latter expressions allow to compute the energy of the system, $\bra{\Psi(\phi_n)}\hat{H}\ket{\Psi(\phi_n)}$, on the  ground-state variational ansatz, $\ket{\Psi(\phi_n)}$, and accordingly to find the value of the parameter, $\phi^*_n$, yielding the minimum of the energy,
\begin{equation}\label{eq:tanphi}
\tanh 2\phi^*_n = -  \frac{2 \lambda (2 n +1)}{(\epsilon_+ + \epsilon_-) }.
\end{equation}
Using this equation we can, for instance, evaluate the  population (and the coherences) of the $|+\rangle$ level on the ground state $|\Psi(\phi^*_n)\rangle$
\begin{equation}\label{eq:pop}\begin{split}
\langle a_+^\dag a_+ \rangle &= \sinh^2 \phi^*_n =\frac{1-\sqrt{1-\frac{4 \lambda^2 \left(2 n+1\right){}^2}{\left(\epsilon _-+\epsilon _+\right){}^2}}}{2 \sqrt{1-\frac{4 \lambda^2 \left(2 n+1\right){}^2}{\left(\epsilon _-+\epsilon _+\right){}^2}}},\\
\av{a_+ a_-} &=\cosh \phi^*_n \sinh \phi^*_n = -\frac{\lambda \left(2 n+1\right)}{ \left(\epsilon _-+\epsilon _+\right) \sqrt{1-\frac{4 \lambda^2 \left(2 n+1\right){}^2}{\left(\epsilon _-+\epsilon _+\right){}^2}}}.
\end{split}\end{equation}
From Eq.~\eqref{eq:tanphi}, it follows that a real solution exists if  
\begin{equation}\label{condition}
\abs{\frac{2 \lambda (2 n +1)}{(\epsilon_+ + \epsilon_-) }} <1;
\end{equation} 
for parameters  not satisfying this relation,  the system exhibits an unstable behavior. 
The physical interpretation of Eq.~\eqref{condition} is that the strength of the photon-mediated interaction must be smaller than the quadratic Zeeman and AC Stark shifts for the system to be stable.

Indeed, a stability condition akin to~\eqref{condition} was already recognized in the context of coherent dissociation of a molecular condensate into a multiple-mode atomic one~\cite{vardi, yuro, anglin, kayali}; if the molecular mode is highly occupied, one can linearize the Hamiltonian around the latter, and describe the process of dissociation with an Hamiltonian formally equivalent to~\eqref{eq:ham} (in our system, the role of the highly occupied mode is taken by the level $|0\rangle$). As a result of this, the  coupling term  $\propto \lambda$ does not conserve the number of particles created (annihilated) by $a_{\pm}$ ($a^\dag_{\pm}$) and for couplings violating~\eqref{condition}, the eigenvalues of~\eqref{eq:ham} becomes complex~\cite{vardi, yuro, anglin, kayali} signalling an unstable character of the dynamics  (cf. with Eq.~\eqref{hamdiag} below).
%
%
%
%
%
%
\begin{figure}[t!]
\includegraphics[width=8.6cm]{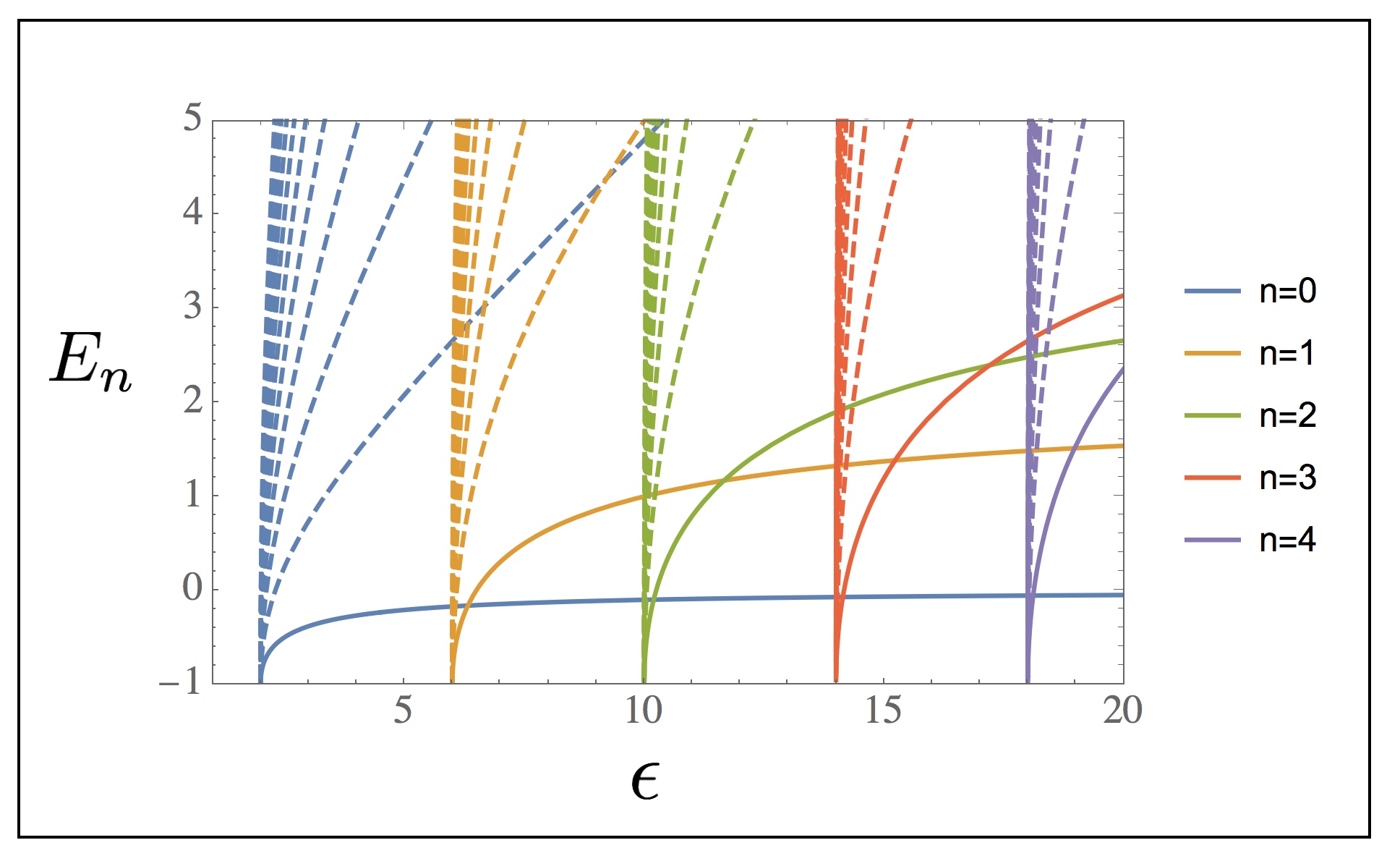}
\caption{\emph{Photons' staircase}:  the ground state energy in each of the sectors with a fixed number of photons, $n$, is one of the  steps of an energy staircase, plotted as a function of $\epsilon$.   Colours denote different bosonic occupation numbers ($\lambda=1$ and $\omega=2$ in the plot). The dotted lines are the energies of the first few excited states in each one of the manifolds with different values of $n$.} 
\label{fig:diag1}
\end{figure}

The procedure resulting from Eq.~\eqref{eq:WF}, is equivalent to diagonalize the Hamiltonian~\eqref{eq:ham}  through the Bogolyubov rotation
\begin{equation}\label{eq:bogo}
a^\dag_+=u_nd^\dag_{1,n}+v_nd_{2,n},\quad a^\dag_-=u_nd^\dag_{2,n}+v_nd_{1,n},
\end{equation}
with $u_n=\cosh\phi^*_n$ and $v_n=\sinh\phi^*_n$; the angle $\phi^*_n$ is, as usual, determined by requiring that off-diagonal terms proportional, for instance, to $d_{1,n}d_{2,n}$ and its hermitian conjugate vanish (the result coincides with Eq.~\eqref{condition}).%
~We can therefore  write the diagonal form of~\eqref{eq:ham} in a sector with fixed number of photons, $n$,
\begin{equation}\label{hamdiag}\begin{split}
\hat{H}=E_0+
&\frac{1}{2}\sum_{m=1,2}\Big((-1)^{m}{(\epsilon_--\epsilon_+)}+\\
&+{\epsilon}\sqrt{1-\frac{4\lambda^2(2n+1)^2}{\epsilon^2}}\Big)d_{m,n}^\dag d_{m,n},
\end{split}\end{equation}
where  the ground state energy of the system reads
\begin{equation}
E_0=\omega n+\frac{1}{2} \epsilon \left(\sqrt{1-\frac{4 \lambda^2 \left(2 n+1\right){}^2}{\epsilon{}^2}}-1\right).
\end{equation}
The energy $E_0$ draws (as a function of $\epsilon\equiv\epsilon_++\epsilon_-$) a staircase in which each step is associated to  a different value of $n$. This is plotted in Fig.~\ref{fig:diag1} together with few excited states energies (computed from Eq.~\eqref{hamdiag})  plotted as dashed lines.\\

\begin{figure*}[t!] 
   \centering
   \includegraphics[width=6.8in]{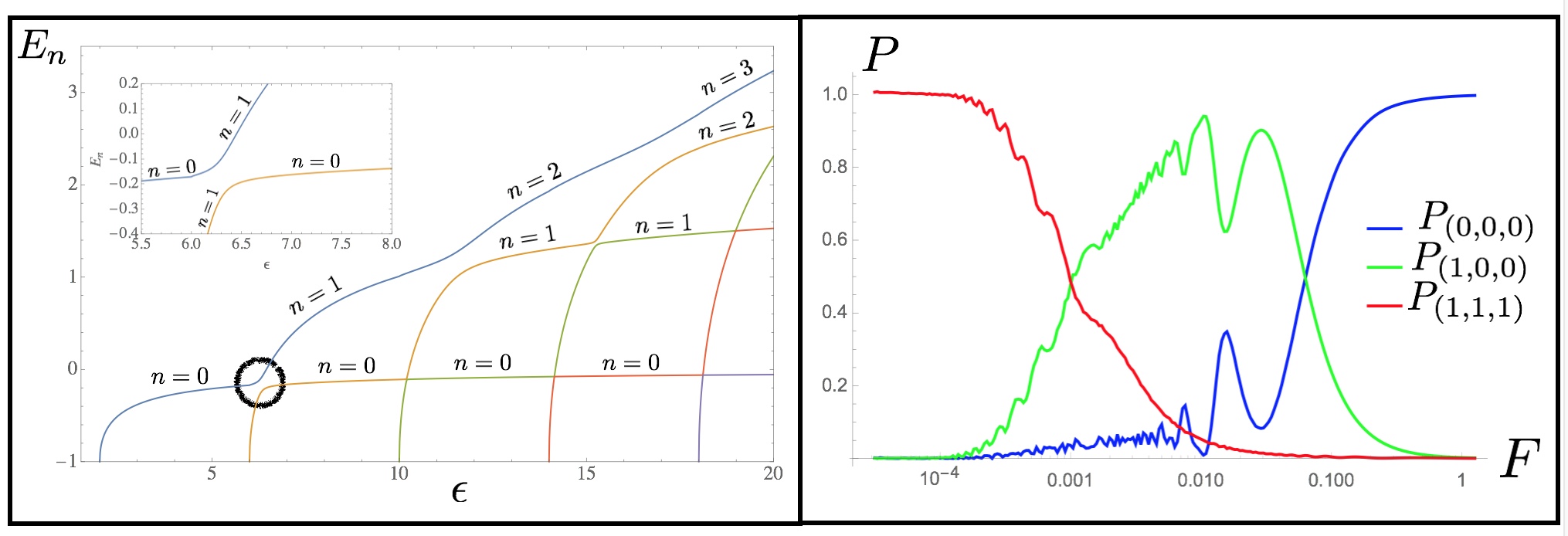}
   \caption{Left panel: energy level repulsion at weak photon pumping $\Omega=0.1$; different photon numbers are indicated over the energy curves of the staircase as a function of $\epsilon$. The parameters $\lambda=1$, $\omega=2$ are the same as in Fig.~\ref{fig:diag1}. Similar staircase structures are observed in the current-voltage characteristic of the Coulomb blockade~\cite{Alha}. Inset: zoom of the avoided crossing between the energies of the states $|\Psi_0(\phi)\rangle$ and $|\Psi_1(\phi)\rangle$ as a function of $\epsilon$. Right panel:  probability (as a function of the ramp speed, $F$) to remain in the ground state with zero photons (blue line; $P_{n=0, m_1=0, m_2=0}$), to transit into the ground state with one photon (green line; $P_{n=1, m_1=0, m_2=0}$), to transit into the first excited state of the manifold with one photon (red line; $P_{n=1, m_1=1, m_2=1}$), starting from the ground state of the manifold with zero photons. There exists an intermediate window of ramp speeds, $F\simeq10^{-3}\lesssim F^*\lesssim10^{-1}$, where the transition occur between ground states, without involving higher excited ones (we have checked that this scenario remains basically unaltered when we add the next excited state in our analysis, see Fig.~\ref{figapp} in the Appendix).  When the drive is too fast (large $F$), the system remains instead frozen in the ground state with zero photons, as expected.} 
   \label{fig_2}
\end{figure*}
%

\section{Landau-Zener theory\\ of the photons' staircase}

The staircase structure facilitates the preparation of a desired number of photons.
In order to illustrate this aspect, we add to the hamiltonian, ${H}$, a term, $V=\Omega(b+b^\dag)$, accounting for  coherent pumping of photons into the system at rate $\Omega$\begin{equation}
{H}' = {H} + \Omega (b+b^\dag).
\label{eqpump}\end{equation}
Corrections to the spectrum $\propto\Omega$ are plotted in Fig.~\ref{fig_2}, and they can be evaluated exactly diagonalizing ${H}'$ for few energy levels, using as basis  the eigenstates of the unperturbed Hamiltonian $\ket{\varphi_n}=  \ket{\Psi_n}\ket{n}$ (from now on we have dropped the dependence from $\phi^*_n$  in $|\Psi(\phi^*_n)\rangle$ to lighten the notation). Our goal is to study Landau-Zener (LZ) transitions among ground state levels with different number of photons, as induced by a time-dependent control parameter, $\epsilon(t)=Ft$.  It is possible to adiabatically climb the staircase without involving excited states in these transitions, if the drive $\epsilon(t)$ occurs at an intermediate speed $F^*$  (see detailed discussion below).
%
%
%
%

%

\begin{figure}[!b]
\centering
\includegraphics[width=5cm]{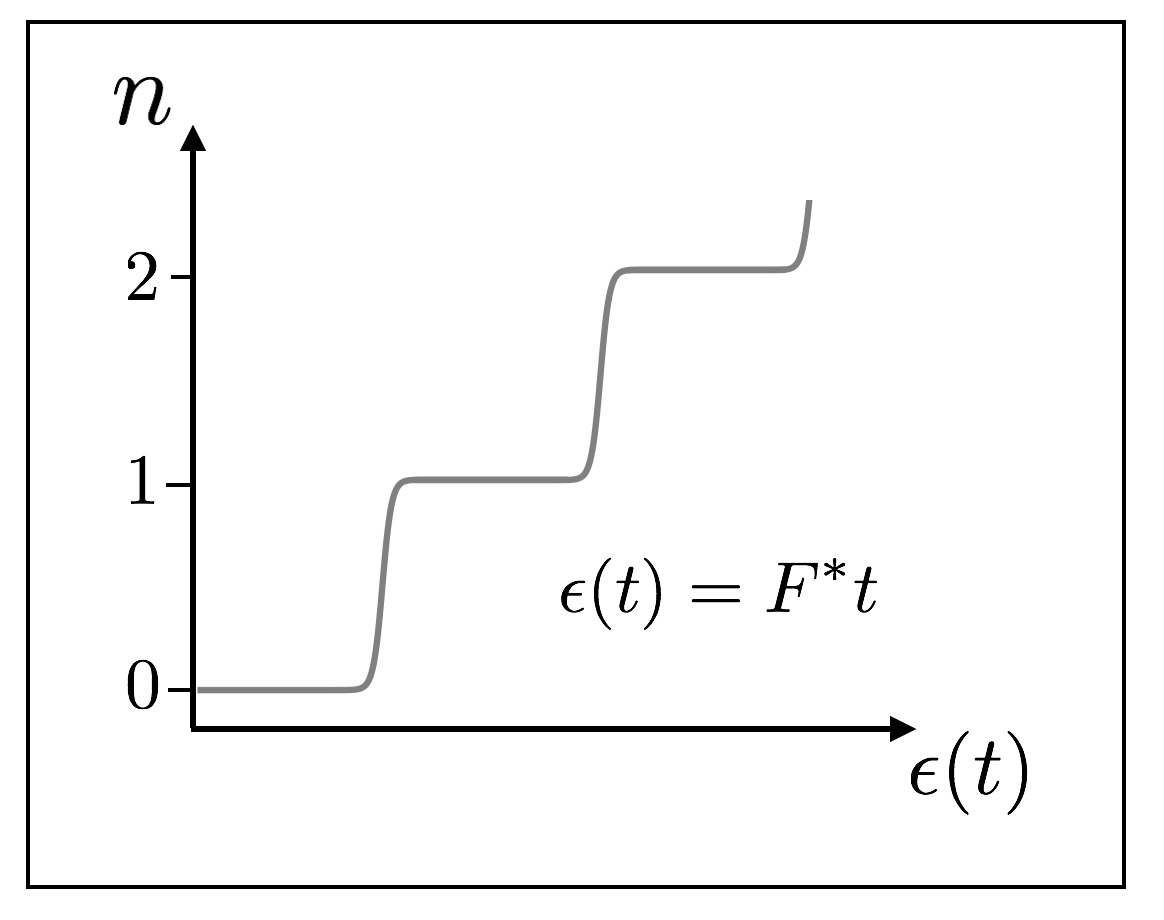}
\caption{ The number of photons increases by a quantised value, as $\epsilon(t)$ is driven through an energy level splitting with the optimal speed $F^*$, which allows direct transitions  between ground states of adjacent photonic manifolds. }
\label{figcb}
\end{figure}
We start writing the Hamiltonian in the basis $|\varphi_n\rangle$, where it looks tri-diagonal since the perturbation   couples   states  differing only by one photon; these are represented by   off-diagonal terms   in the following matrix representation of  ${H}'$
\begin{equation}\label{eq:HamMat}
\begin{split}
&\bra{\varphi_n}{H}' \ket{\varphi_m} =\sum_{n} E_n  
+ \sum_{n,m} (  \Omega \sqrt{n+1} \langle{\Psi_{n}}| {\Psi_{m}}\rangle\delta_{m,n+1}+\\
&+\Omega \sqrt{n}  \langle{\Psi_n}|   {\Psi_{m}} \rangle \delta_{m,n-1}).
\end{split}\end{equation}
The overlap $\langle{\Psi_n}|   \Psi_{n\pm1}\rangle$ is calculated following Ref.~\cite{perel}  as (see also Appendix)
\begin{equation}\label{grounds}\begin{split}
\langle\Psi_n| \Psi_{n\pm1}\rangle&= \av{e^{-(\phi_n  a_+^\dag a_-^\dag + \textsl{h.c.})} e^{(\phi_{n\pm1}  a_+^\dag a_-^\dag - \textsl{h.c.})} }=\\
&=\frac{1}{  \cosh\left( \phi_{n\pm1} - \phi_{n}\right)}.
\end{split}\end{equation}
In order to gain intuition for the perturbative corrections induced by a weak photon pumping on the photon staircase spectrum, we first consider a simple perturbative analysis in the parameters' regime $\Omega$, $\lambda\ll \epsilon$.
In this limit, the overlap reads $\langle{\varphi_n}|\hat{b} |  {\varphi_{n+1}} \rangle \approx\sqrt{n}  \left( 1 - {2\lambda^2}/{\epsilon^2}\right)$, and the ground  state energies,  $E_n=\omega n-{\lambda^2}/{\varepsilon}(2n+1)^2$.
Let us now consider an energy level $E_n$, with photon number $n$, crossing with a level with energy $E_{n'}$ and $n'=n+1$; at their intersection, occurring at $\varepsilon\simeq\varepsilon^*_{n,n'}$, a straightforward application of degenerate perturbation theory  in $\Omega$, yields an energy splitting 
\begin{equation}
\Delta E_{n,n'} \simeq
2\Omega~\sqrt{n+1}\left(1-\frac{1}{32(n+1)^2}\frac{\omega^2}{\lambda^2}\right).
\end{equation} 
Beyond this simple analysis, an exact numerical evaluation of the eigenvalues of the energy matrix~\eqref{eq:HamMat}, as a function of $\epsilon$, provides the energy level structure portrayed in the left panel of Fig.~\ref{fig_2}. A small pumping rate,   $\Omega$, is sufficient to induce an effective energy level repulsion reshaping the staircase structure into a sequence of avoided crossings between ground states with different photon numbers. We remark that, although the coherent photon pumping does not commute with the unperturbed Hamiltonian~\eqref{eq:ham}, $[{V},{H}]\neq0$,  its effect, for small $\Omega$, is negligible for values of $\epsilon$ away from the  crossing points $\varepsilon^*_{n,n'}$, and in these regions we can still effectively consider $\hat{n}_b$  a good quantum number.

According to this structure, an adiabatic  climbing of the staircase from  a state with zero photons (blue line in the left panel of Fig.~\ref{fig_2}) to a state with a certain photonic population, can  be designed as follows: We start from the ground state with zero photons and given initial $\epsilon_0$ at time $t_0=0$, and we  drive linearly in time the control parameter $\epsilon(t)=Ft$, with  rate $F$; following the argument presented above, the drive can induce a transition to the ground state with one photon as $\epsilon(t)$ approaches the crossing point located at $\epsilon\simeq\varepsilon_{0,1}$ (see the zoom  of the crossing among $E_0$ and $E_1$ in the region $5.5<\epsilon<7$: inset of left panel of Fig.~\ref{fig_2}). 

For this LZ analysis, transitions involving excited states do not play a significant role. For instance, the  transition from a ground state with $n$ photons into the first excited states of the next photonic manifold, such as  a transition from $|\Psi_{n}\rangle$ to $d^\dag_{1,n+1}d^\dag_{2,n+1}|\Psi_{n+1}\rangle$, or to $(d^\dag_{1,n+1})^2(d^\dag_{2,n+1})^2|\Psi_{n+1}\rangle$, are assisted  by  matrix elements of the perturbation $H'$, which are smaller than the one connecting the two ground states. This is shown in the Appendix: Eqs.~\eqref{primoover} and \eqref{secondover} display the matrix elements for the transition between the ground state and these two excited ones, and they should be compared with Eq.~\eqref{grounds}, reporting the overlap between ground states. The overlaps involving excited states are always smaller by a factor proportional to increasing powers of $\tanh(\phi_{n+1}-\phi_n)$ (a quantity always smaller than one) as higher excited states are considered. In particular, as  $\epsilon$ grows large (i.e. $\epsilon(t)$ increases, while the photonic manifolds of the staircase are  explored)  these overlaps are suppressed  algrebraically in $1/\epsilon$. We have  numerically explored  the specific case of the transition between the ground states of the first two photonic manifolds ($n=0$ and $n=1$) in the right panel of Fig.~\ref{fig_2} including the first  excited state, and found that the weaker coupling to excited states discussed above, results into the possibility to perform adiabatic transitions among ground states. In  Fig. 6 of the Appendix we show that the quantitative features of this adiabatic transition remain unaltered upon inclusion of the next excited state. 

The right panel of Fig.~\ref{fig_2} shows that a slow ramp would favour the transition to the first excited state of the manifold with one photon, but at intermediate ramp speeds, instead, the probability to transit into the ground state $|\Psi_1\rangle$ is the dominant one. As  $\epsilon(t)$ increases further, the subsequent transitions will basically occur between ground states of manifolds with different photon numbers (if the ramp is moderately slow), since, as discussed in the paragraph above, for $\epsilon\gg1$ the  matrix elements of the operator controlling the transition, $H'$,  become parametrically smaller than those connecting ground states $|\varphi_n\rangle$ and  $|\varphi_{n+1}\rangle$ (see also the explicit expressions of these overlaps in the Appendix, Eqs. (A5) and (A6)).
Therefore, a sequence of LZ-like transitions allows to climb the staircase and to achieve a target number of photons.


%

We observe that the jumps between the steps of the staircase characterised by different integer values of photons (see Fig.~\ref{figcb}), and explored as  $\epsilon(t)$ is increased in time, recalls the current-voltage staircase profile observed in the phenomenon of Coulomb blockade~\cite{Alha}.
Although the underlying mechanism is different, the two cases share the feature that every step of the staircase corresponds to a state with distinctly resolved physical properties: in our quantum optics set-up, for intervals of $\epsilon$ away from avoided crossings, each step of the staircase is associated to a fixed and quantised number of photons with negligible fluctuations.\\
%
%

Naturally,  in the case of LZ photon preparation we expect that dissipation will classicalise the state of light at late times, but the staircase structure of the photonic response (see Fig~\ref{figcb}) guarantees that, at intermediate times, the light degree of freedom will be found in a quantum state with a well defined number of photons and few fluctuations on the top of them, provided the time $t_f$ to implement the LZ ramp satisfies the condition $t_f\ll 1/\kappa$.  Furthermore, in order to have an adiabatic  ramp  and to remain in a weak photon pumping regime, we require also that $t_f > 1/\lambda$.  Since the coupling strength $\lambda$ depends on the power in the optical drive fields and on detunings from atomic and cavity resonance, the ultimate limits are set by atomic and cavity parameters.

In particular, climbing the staircase requires large collective cooperativity $\mathcal{N}_0 C$, where $C = 4g_\pm^2/(\kappa \Gamma)$ is the single-atom cooperativity given the atomic excited-state linewidth $\Gamma$.  This requirement is derived from the scaling with $\lambda$ of the atomic spontaneous emission rate $\Gamma_\mathrm{sc} \sim \lambda\delta/(\mathcal{N}_0 C\kappa)$ and of Raman scattering into the cavity at rate $\gamma \sim \lambda \kappa/\delta$.  At an optimal detuning $\delta \sim \sqrt{\mathcal{N}_0 C}\kappa$ from Raman resonance, the coupling-to-dissipation ratio scales as $\lambda/(\gamma+\Gamma) \sim \sqrt{\mathcal{N}_0 C}$.  Thus, the requirement $\lambda > \kappa$ for climbing the staircase can be satisfied for large collective cooperativity $\mathcal{N}_0 C \gg 1$.  Collective cooperativities $\mathcal{N}_0C > 10^4$ are routinely achieved with atomic ensembles in optical cavities, making the staircase accessible to current experiments.

\section{Driven-dissipative dynamics} 
We now consider the competition between coherent pumping and photon  losses, occurring at rate $\kappa$ and described by the jump operator $L=b$ (photon losses occurring during the intermediate processes contributing to the two-body resonance described in Fig.~\ref{fig1} are negligible in the far resonance regime $\delta\gg1$ as discussed in Ref.~\cite{monikanew}). 
In particular, we will  explore dynamics for times $t\gg1/\kappa$, where  light becomes Poissonian as result of decoherence, at variance with the conditions discussed at the end of the previous Section (we will still  assume large cooperativity and $\delta\gg1$, though).

We consider a sudden switch  of coherent pumping at times $t>0$ to counterbalance cavity losses. 
We  prepare the system in the ground state of~\eqref{eq:ham} with zero photons, $|\Psi(\phi^*_0)\rangle$, and we consider the time evolution ruled by the following set of equations of motion for the  expectation values of the 'molecular' degrees of freedom:  atomic coherences, $\mathcal{C}_{\pm}\equiv a^\dag_+ a^\dag_-\pm a_-a_+$, and populations, $\mathcal{P}\equiv  a^\dag_+ a_++a^\dag_-a_-$, 
\begin{widetext}
\begin{equation}\begin{split}
\label{sist1}
i\frac{d{\langle\mathcal{C}}_-\rangle}{dt}&=-\epsilon\langle\mathcal{C}_+\rangle-2\lambda(2\langle n\rangle+2\langle\mathcal{P}\rangle+2\langle n\rangle\langle\mathcal{P}\rangle+1)
-4\lambda(\langle b^\dag\mathcal{P}\rangle_c\langle b\rangle+\langle b\mathcal{P}\rangle_c\langle b^\dag\rangle),\quad i\frac{d\langle{\mathcal{C}}_+\rangle}{dt}=-\epsilon\langle\mathcal{C}_-\rangle,\\
i\frac{d\langle{\mathcal{P}}\rangle}{dt}&=4\lambda\langle\mathcal{C}_-\rangle(1+\langle n \rangle)+4\lambda(\langle b^\dag\mathcal{C}_-\rangle_c\langle b\rangle+\langle b\mathcal{C}_-\rangle_c\langle b^\dag\rangle),\\
\end{split}
\end{equation}
coupled to the dynamics of photons
\begin{equation}\begin{split}
\label{sist2}
\frac{d\langle{b}\rangle}{dt}&=-i\Omega-\frac{\kappa}{2} \langle b\rangle-i\omega \langle b\rangle-2i \lambda \langle \mathcal{C}_+ \rangle \langle b \rangle  -2i \lambda \langle b \mathcal{C}_+\rangle_c , \quad \frac{d\langle{n}\rangle}{dt}=i\Omega(\langle b\rangle-\langle b^\dag\rangle)-\kappa \langle n\rangle,\\
\frac{d\langle b^2 \rangle_c}{dt}&=-2i\omega\langle b^2 \rangle_c-\kappa\langle b^2\rangle_c -4i\lambda\langle \mathcal{C}_+\rangle\langle b^2\rangle_c-4i\lambda\langle{b} \rangle \langle b \mathcal{C}_+ \rangle_c  ,
\end{split}\end{equation}
where we have assumed the atomic and photonic degrees of freedom to be in a Gaussian state, and we have included terms describing three-body correlations between light and matter degrees of freedom, such as $\langle b \mathcal{C}_+\rangle_c\equiv  \langle b \mathcal{C}_+\rangle-\langle b\rangle \langle \mathcal{C}_+\rangle$. The equation of motion for the the latter reads
\begin{equation}\label{corrng}
 \frac{d\langle b\mathcal{C}_+\rangle_c}{dt}=-(\kappa/2+i\omega)\langle b\mathcal{C}_+\rangle_c+i\epsilon\langle b\mathcal{C}_-\rangle_c
 -2i\lambda\langle \mathcal{C}_+\rangle(\langle b\mathcal{C}_+\rangle_c-\langle b\rangle\langle \mathcal{C}_+\rangle).
\end{equation}
\end{widetext}

\begin{figure}[h!]
\centering
\includegraphics[width=8cm]{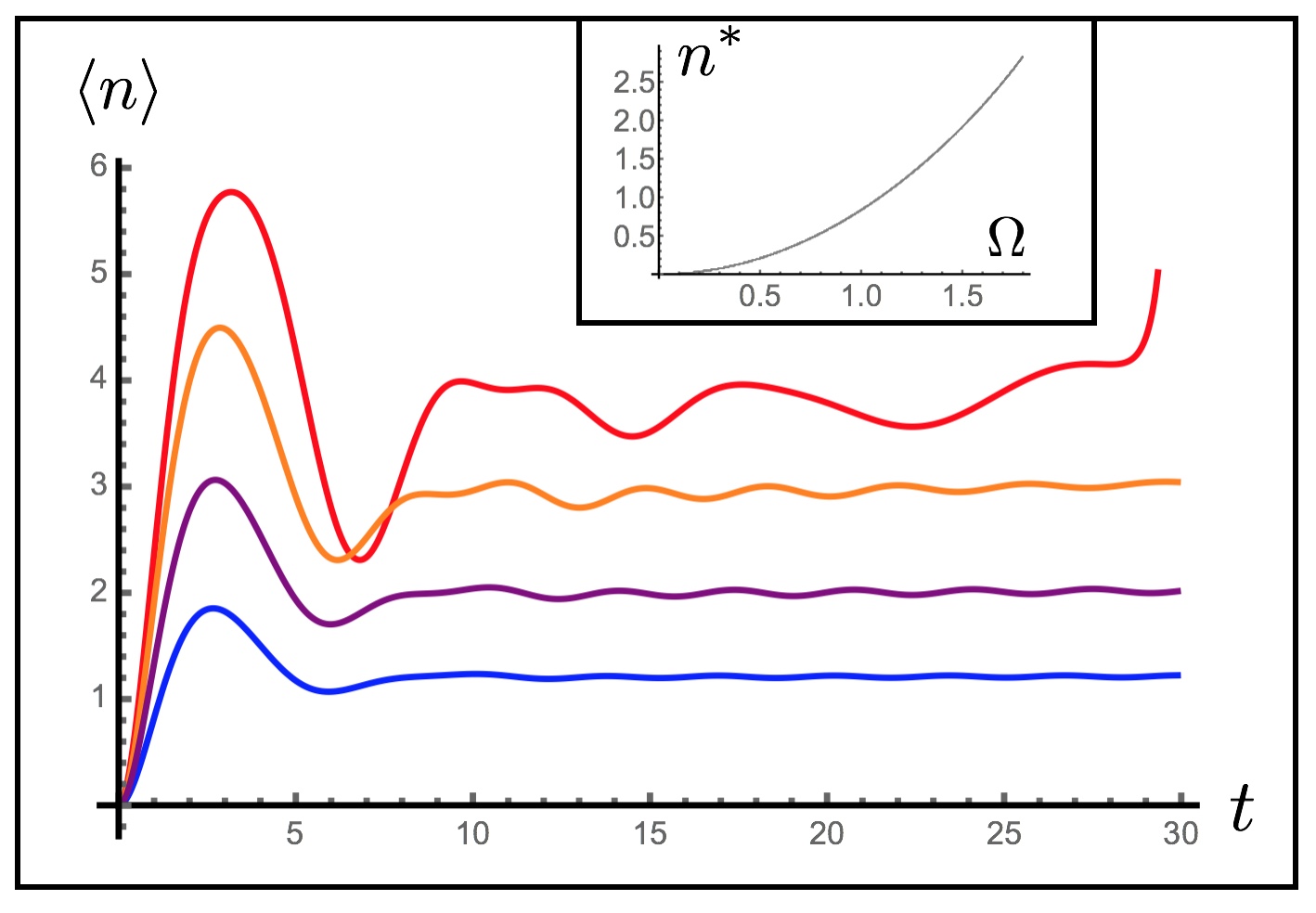}
\caption{ Dynamics of the photon number, $\langle n \rangle$, for increasing values of $\Omega$. The blue ($\Omega=1.2$), purple ($\Omega=1.53$), orange ($\Omega=1.83$), curves correspond to the formation of $n^*\simeq1, 2, 3$ photons respectively (in this figure: $\epsilon=10$, $\omega=1$, $\lambda=0.1$, $\kappa=1$). When $\Omega\simeq2.03$ (red curve), the departure from the metastable photonic plateau (with $n^*\simeq4$) occurs on time scales which  can be resolved. Inset: time average, ${n}^*$, of  $\langle n \rangle$ in the plateau as a function of the photonic pumping, $\Omega$. The fit is parabolic, ${n}^*=\alpha \Omega^2$, with $\alpha=0.85$.} 
\label{fig4}
\end{figure}

%
%

During driven-dissipative dynamics the magnitude of three-body correlations between light and matter in Eqs.~\eqref{sist1} and~\eqref{sist2} remains small, with the consequence that the two sectors are almost dynamically decoupled. This 
allows the number of photons to relax towards a steady state value, $n^*$, while the  atomic excitations, $\langle \mathcal{P}\rangle$, are still slowly growing as a consequence of the pumping. The steady-state number of  photons generated is  predicted by the   formula
\begin{equation}
\label{nasy}
n^*=\frac{\Omega}{\kappa}\frac{\kappa(\Omega+2\lambda\operatorname{Re}\langle b\mathcal{C}_+\rangle^*)+4\lambda\omega\operatorname{Im}\langle b\mathcal{C}_+\rangle^*}{(\kappa/2)^2+\omega^2},
\end{equation}
where $\langle b\mathcal{C}_+\rangle^*$ is the asymptotic steady state value of $\langle b\mathcal{C}_+\rangle$.
The time-scales separation in the dynamics of light and matter degrees of freedom is at the origin of the   metastability of the plateau reached at long times by $\langle n \rangle$. In fact, a slow  growth of $\langle P\rangle$ (occurring while the photons' steady state is already established), provokes a growth in $\langle C_+\rangle$ as well, which acts as a source in Eq.~\eqref{corrng} ruling the dynamics of $\langle bC_+\rangle$. 
%
%
%
Therefore, the combination of  bare photon pumping controlled by $\Omega$, and of  effective pumping induced by three-body light-matter correlations,   determines at later times an increase of $\langle n \rangle$, which can drive the system beyond the  parameters' region delimited by the  condition in Eq.~\eqref{condition}, where eigenenergies become imaginary (the associated critical value of  the interaction strength $\lambda$  in~\eqref{condition}, is renormalised to a lower value after the inclusion of three-body light-matter correlations).

At the timescales for the departure from the metastable steady state, also the atomic and photonic correlations, evolving respectively under Eqs.~\eqref{sist1} and~\eqref{sist2}, experience a quick, diverging growth.
The characteristic time, $t_{b}$, for the  departure from the metastable photonic steady-state is  proportional to  $\Omega$:
in Fig.~\ref{fig4} we portrait time-resolved profiles of $\langle n\rangle$ at increasing pumping rates, $\Omega$, and we illustrate  the metastability of the dynamics of $\langle n \rangle $ as $\Omega$ becomes sufficiently large.
Fitting the breakdown time $t_b$ of the photonic plateau one finds that $t_b\propto \epsilon^2/\lambda$, which can be controlled both via the quadratic Zeeman and AC Stark shifts, and the occupation of the mode $|0\rangle$, as it can be realised by inspection of Eqs.~\eqref{quanti} .

The inset of Fig.~\ref{fig4}, displaying the average number of photons as function of $\Omega$, demonstrates instead that the quantization of $n$ typical of the staircase structure (still present when the photonic pumping is adiabatically switched and photons generated via slow LZ transitions) is lost when pumping and  dissipation are suddenly turned on, since ground and excited states of the staircase are strongly mixed in this case.
%
%
%
%

Photon generation using ramps and LZ transitions are indeed more efficient than the sudden switching of photon pumping:
In each one of the plateaux of Fig.~\ref{figcb} the photonic degree of freedom is in a state with fixed and quantised number of photons, $n$,  provided dissipation, $\kappa$, is weak enough to affect the dynamics of the system only at late times (see discussion at the end of Sec.~IV).
On the contrary, suddenly switching the photon pumping results in a transient dynamics with no quantised photon number (cf.~Fig.~\ref{fig4}), which asymptotes  to a plateau  where quantum features have been erased. 
Specifically, we have resolved the dynamics of our model combining a Gaussian ansatz for the atoms,  as done in Eqs.~\eqref{sist1}, with a truncated ansatz for the density matrix of the light, $\rho\equiv{\sum}_{n,m}\rho_{nm}|n\rangle\langle m|$, with $n,m=1, ...M$ and $M=11$ (see Appendix).
Although in the plateaux shown in Fig.~\ref{fig4} the presence of the atomic degrees of freedom sizeably enhances the asymptotic expectation value of $\langle n\rangle$ compared to the decoupled ($\lambda=0$) case (and therefore the system is in a state where light and matter are hybridised), the light generated is classical, as we have checked by calculating  the photonic variance, $\Delta n^2=\langle n^2\rangle-\langle n \rangle^2$, from the density matrix ansatz, $\rho$, found always very close to $\langle n \rangle$ -- a signature of the classical nature of the light produced in the cavity.  
This occurs for times $t\gg 1/\kappa$, when the system has reached the steady state and at the same time the dissipation has washed  out any quantum feature  present at short times (see Appendix).

\section{Perspectives }
~As a future direction, it would be interesting to study  a many-body version of the problem analysed in this work, which can be realised, for instance, considering a one-dimensional lattice of several cavities (modelled as in Fig.~\ref{fig1}), connected one to each other by next-neighbour photonic hoppings (in the spirit of an Hubbard model; see for similar ideas in quantum optics the review in~\cite{houck12}). 
Studying the competition of this kinetic term with the 
 driving and  dissipation discussed in this work, would pave the way to a quantum many body  simulator for the preparation of multi-photon states, which would benefit of the  tunability properties of the photons' staircase as a leverage for  experimental implementations. It would be, for instance, intriguing to look for driven-dissipative phase transitions in this many-body version of our system following the directions mentioned in the introduction.\\

\section{Acknowledgments }  We acknowledge  discussions with F. Reiter. JM acknowledges support by the EU Horizon 2020 research and innovation program under Marie Sklodowska-Curie Grant Agreement No. 745608 (MC).
YS and ED acknowledge support from 
Harvard-MIT CUA, NSF Grant No. DMR-1308435,
AFOSR-MURI Quantum Phases of Matter (grant FA9550-14-1-0035),
AFOSR-MURI Photonic Quantum Matter (award FA95501610323).

\bibliography{biblio}

\begin{appendix}
\begin{widetext}

\section{Matrix elements for the Landau-Zener transition}

In this section of the Appendix we detail the calculation of the matrix elements of the perturbation $V$ connecting ground and  excited states of the staircase, involved in the  study of the transitions in Fig.~\ref{fig_2}.
~The squeezed ground state~\eqref{eq:WF}  can be  represented in the Fock basis of the occupation numbers of  the modes, $a_\pm$, as (see for instance Ref.~\cite{perel})
\begin{equation}\label{eq:A1}
\ket{\Psi_n} = \frac{1}{\cosh \phi_n} \sum_l (\tanh \phi_n)^{l} \ket{l,l}, 
\end{equation}
where $\phi_n$ is the squeezing angle for a fixed number of photons, $n$. We describe the excited states of the system using the quasiparticle creation operators $d^\dag_{1,n}$ and $d^\dag_{2,n}$ introduced in Eq.~\eqref{eq:bogo}. We represent these operators inverting the Bogolyubov rotation~\eqref{eq:bogo}:
\begin{equation}\label{eq:qpop}
d^\dag_{1,n} = u_n a^\dag_+ - v_n a_-, \quad d^\dag_{2,n} = - v_n  a_+ + u_n a_-^\dag.
\end{equation}
Any excited state of the system can be represented as
\begin{equation}
\ket{k, k', n} = ( d_{1,n}^\dag  )^k (  d_{2,n}^\dag )^{k'} \ket{\phi_{n}},
\end{equation}
with $k>0$ and $k'>0$.

In Eq.~\eqref{eqpump} we introduced a term accounting for coherent pumping of photons into the system at rate $\Omega$, $\hat V = \Omega (  b^\dag +  b)$. This term  introduces mixing between the ground state in the sector with $n$ photons and excited states in the sector with $n+1$ photons. In order to account for this effect, we calculate the overlap of the excited states in the sector with $n+1$ photons with the ground state of the manifold with $n$ photons, $\ket{\phi_{n}}$.

First of all, we notice that the overlap between the excited states with $k$ quasiparticle excitations of only  one  type, and the ground state in the neighbouring sector, is equal to zero for any number of excitations ($k>0$):
\begin{eqnarray}
\Omega\bra{0,0,n+1}&&(  b^\dag +  b)\ket{k,0,n}  =\Omega \sqrt{n+1}\bra{0,0,n+1} (  d_{1,n})^k \ket{0,0,n} = 0.
\end{eqnarray}
This is a consequence of the fact that the ground state~\eqref{eq:A1} can be written as a superposition of states with the same number of excitations in the ($\pm$) sectors of  the Fock space of the original degrees of freedom of the model. 

The non-zero overlaps with  excited states induced by the perturbation operator $ V$ is between the state with $k$ quasiparticles in both  ($\pm$)  modes, $\Omega\bra{0,0,n+1}(  b^\dag +  b)\ket{k,k,n}$. The first non-trivial overlap is $ V_{1,1,1} \equiv \Omega\langle 0,0,0|1,1,1\rangle$.
We calculate this overlap using  the representation~\eqref{eq:A1},
\begin{equation}\label{primoover}
\begin{split}
&V_{1,1,1} =\Omega\langle 0,0,0|1,1,1\rangle=\Omega \bra{\phi_0} \hat d_{1,(n=1)}^\dag \hat d_{2,(n=1)}^\dag \ket{\phi_1}=\\
&=\frac{\Omega}{\cosh\phi_0\cosh\phi_1}\sum_{l,l'}(\tanh\phi_0)^l(\tanh\phi_1)^{l'}\langle l',l'| (- v_n  a_+ + u_n a_-^\dag)(u_n a^\dag_+ - v_n a_-)  |l,l\rangle=\\
&=\frac{\Omega}{\cosh\phi_0\cosh\phi_1}\sum_{l,l'}(\tanh\phi_0)^l(\tanh\phi_1)^{l'}(-\delta_{l,l'}(2l+1)u_nv_n+\delta_{l',l+1}(l+1)u^2_n+\delta_{l',l-1}lv^2_n)=\\
&=\frac{\Omega}{\cosh\phi_0\cosh\phi_1}(-\cosh\phi_0\sinh\phi_0\sum_{l=0}(2l+1)(\tanh\phi_0)^l(\tanh\phi_1)^{l}+\tanh\phi_1(\cosh\phi_0)^2\sum_l(l+1)(\tanh\phi_0)^l(\tanh\phi_1)^{l}+\\&+\tanh\phi_0(\sinh\phi_0)^2\sum_l(l+1)(\tanh\phi_0)^l(\tanh\phi_1)^{l})=-\Omega\frac{\tanh(\phi_0-\phi_1)}{\cosh(\phi_0-\phi_1)}.
\end{split}
\end{equation}

Analogously, we calculate the overlap 
\begin{equation}\label{secondover}
V_{2,2,1}=\Omega\langle 0,0,0|2,2,1\rangle=\Omega \bra{\phi_0} \left( \hat d_{1,(n=1)}^\dag \right)^2 \left(\hat d_{2,(n=1)}^\dag  \right)^2 \ket{\phi_1}=2\Omega\frac{(\tanh(\phi_0-\phi_1))^2}{\cosh(\phi_0-\phi_1)}.
\end{equation}

\begin{figure*}[t!] 
   \centering
   \includegraphics[width=3.5in]{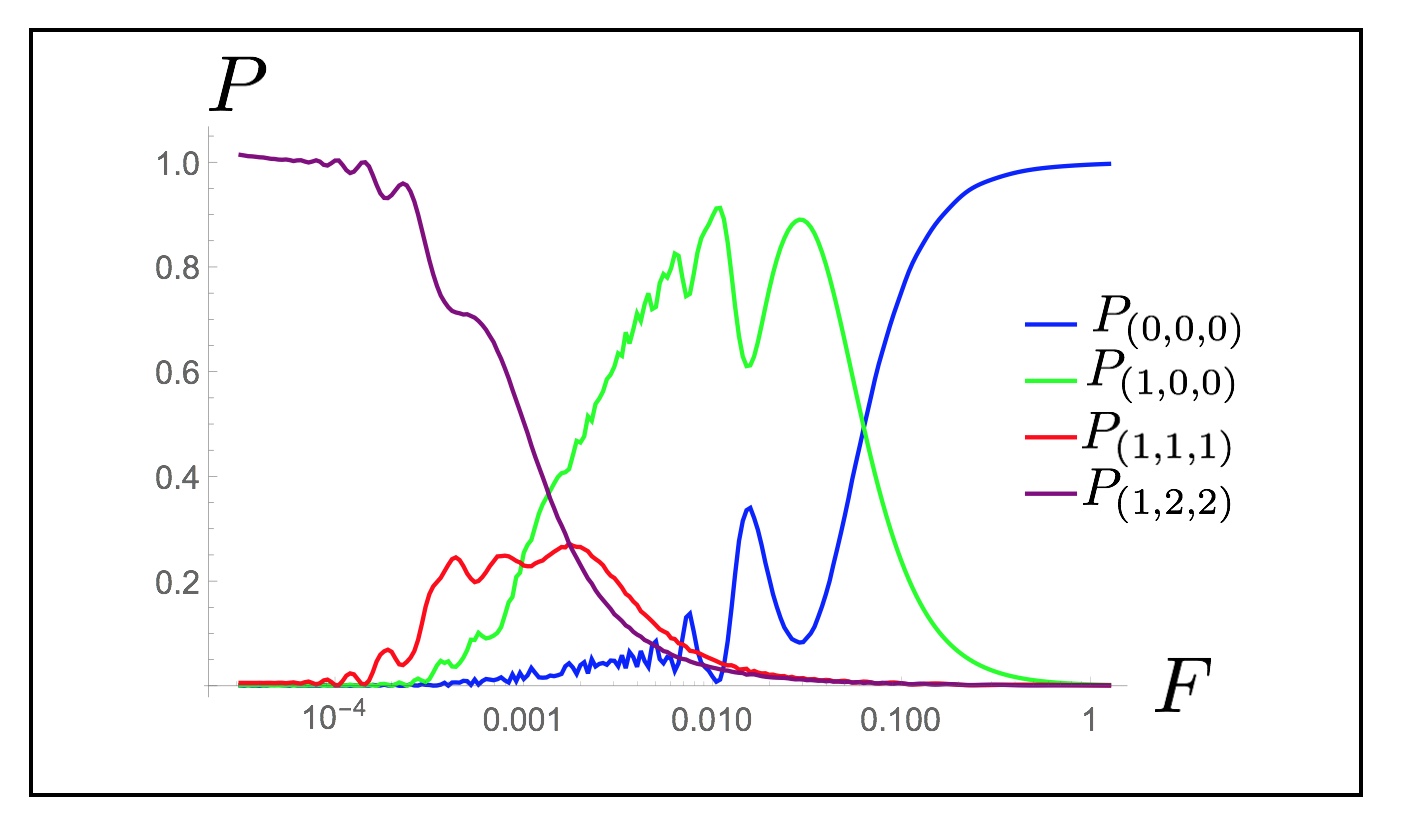}
   \caption{Parameters of the plot: $\lambda=1$, $\omega=2$, $\Omega=0.1$.  Probability (as a function of the ramp speed, $F$) to remain in the ground state with zero photons (blue line; $P_{n=0, m_1=0, m_2=0}$), to transit into the ground state with one photon (green line; $P_{n=1, m_1=0, m_2=0}$), to transit into the first excited state of the manifold with one photon (red line; $P_{n=1, m_1=1, m_2=1}$), to transit  into the second excited state of the manifold with one photon (purple line; $P_{n=1, m_1=2, m_2=2}$) starting from the ground state of the manifold with zero photons. There exists an intermediate window of ramp speeds, $F\simeq10^{-3}\lesssim F^*\lesssim10^{-1}$, where the transition occur between ground states, without involving higher excited ones. When the drive is too fast (large $F$), the system remains instead in the ground state with zero photons, as expected, while at lower speeds, the system transits always into the state $|2,2,1\rangle$.} 
   \label{figapp}
\end{figure*}
In these expressions  one can recognise the overlap between the ground states of adjacent photonic manifolds, given by Eq.~\eqref{grounds}.

These overlaps allow to solve the LZ problem reported in Fig.~\ref{fig_2}b, and to check that adding the next excited state $|2,2,1\rangle$, the feature of an intermediate window of ramp speeds where the transition occurs only involving  ground states, remains substantially unaffected. This is reported  in Fig.~\ref{figapp} of this Appendix. 

From the expressions~\eqref{primoover} and~\eqref{secondover} and the definition~\eqref{eq:tanphi},  one can realise that, at large $\epsilon$, they both become parametrically small, while the overlap between ground states~\eqref{grounds} approaches a constant, as stated in the main text.

\section{Combined Gutzwiller and Gaussian ans\"{a}tze for driven-dissipative dynamics}
In this section of the Appendix we summarise the calculation of the photon variance $\Delta n^2$, for which we resort to the following ansatz for the system density matrix:
\begin{equation}\label{ans}
\rho\simeq\rho^b\otimes\rho^a_G,
\end{equation}
where $\rho^a_G$ is a Gaussian ansatz density matrix for the atomic degrees of freedom, while  for the photonic degree of freedom we write a density matrix in a bosonic Hilbert space truncated up to $M$ bosons: 
\begin{equation}
\rho={\sum}_{n,m}\rho_{nm}|n\rangle\langle m|,
\end{equation} 
with $n,m=1, ...M$ (we use $M=11$ in the following calculations). This is in spirit similar to the Gutzwiller ansatz employed in Ref.~\cite{walt} for the dissipative dynamics of bosons.
Inserting the ansatz \eqref{ans} in the Lindblad equation
\begin{equation}
\dot{\rho}=-i[H,\rho]+\kappa(L\rho L^\dag-\frac{1}{2}\{L^\dag L,\rho \}),
\end{equation}
with $L=b$ incoherent photon losses at rate $\kappa$, we find that
%
the equation of motions for the two-point functions of the atomic degrees of freedom follow Eqs.~\eqref{sist1} (with three-body correlations  set to  zero), with the  difference that 
now $\langle n \rangle=\sum^M_{n=1}\rho_{nn}$, while the  equations~\eqref{sist2} for the one and two-point functions of the photon are replaced by the $M^2-1$ linear system of equations of motion for the matrix elements of $\rho_b$. As initial conditions, we consider $\rho(t=0)=|n\rangle\langle n|\otimes |\Psi_n\rangle\langle \Psi_n|$.
These equations  appear cumbersome for large $M$, but they can be readily derived. Here, for illustrative purposes, we write down   the equations of motion for the population of the $n=0$ mode, $\rho_{00}$, and for the first  coherence, $\rho_{01}$:
\begin{equation}
\begin{split}
\frac{d\rho_{00}}{dt}&=\gamma\rho_{11}-i\Omega(\rho^*_{01}-\rho_{01}),\\
\frac{d\rho_{01}}{dt}&=\gamma(\sqrt{2}\rho_{12}-\frac{\rho_{01}}{2})-i(\Omega\rho_{11}-\sqrt{2}\Omega\rho_{02}-(\omega+2\lambda\langle\mathcal{C}_+\rangle)\rho_{01}-\Omega\rho_{00}).
\end{split}
\end{equation}

As a sanity check we benchmarked our predictions for $\langle n \rangle$ in the  exactly solvable case, $\lambda=0$.\\

The variance is then  straightforwardly  written in terms of the matrix elements of $\rho_b$,
\begin{equation}
\Delta n^2=\langle n^2\rangle-\langle n \rangle^2=\left(\sum^M_{n=1} n^2\rho_{nn}\right)-\left(\sum^M_{n=1} n\rho_{nn}\right)^2.
\end{equation}
An instance of the dynamics of $\Delta n^2-\langle n\rangle$  is reported in Fig.~\ref{figapppp}, showing that at late times, $\Delta n^2\simeq\langle n\rangle$. This circumstance is independent from the specific choice of parameters adopted. 
\begin{figure*}[t!] 
   \centering
   \includegraphics[width=2.5in]{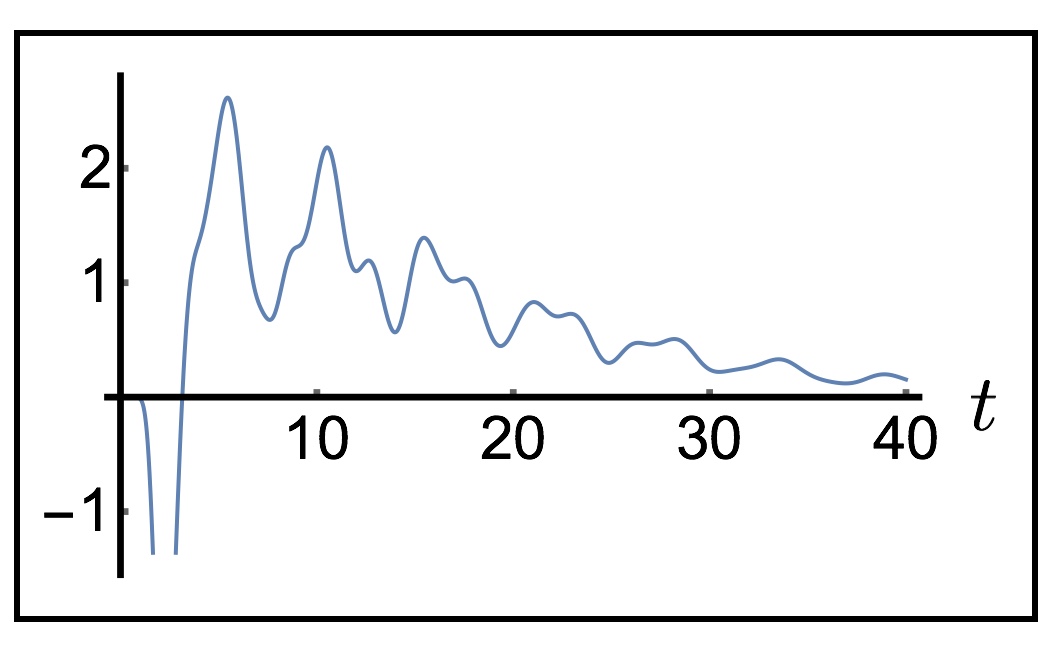}
   \caption{Dynamics of the difference $\Delta n^2-\langle n\rangle$,  for $\Omega=1.53$, $\epsilon=10$, $\omega=1$, $\lambda=0.1$, $\kappa=0.1$, and initial conditions $\rho(t=0)=|0\rangle\langle0|\otimes|\Psi_0\rangle\langle\Psi_0|$.} 
   \label{figapppp}
\end{figure*}

\end{widetext}
\end{appendix}

\end{document}